# Trigger and Data Acquisition for hadron colliders at the Energy Frontier

Wesley H. Smith

*University of Wisconsin - Madison*

## Abstract

The LHC trigger and data acquisition systems will need significant modifications to operate at the HL-LHC. Due to the increased occupancy of each crossing, Level-1 trigger systems would experience degraded performance of the LHC algorithms presently selecting up to 100 kHz of crossings from the LHC input rate of 40 MHz. The DAQ systems will experience larger event sizes due to greater occupancy and higher channel counts of new detectors. This paper summarizes findings and recommendations to upgrade the LHC experiments trigger and data acquisition systems for operation at the HL-HLC.

*To be submitted to the Proceedings of the American Physical Society's Division of Particles and Fields Community Summer Study, 2013.*

Keywords:  Trigger, Data Acquisition, LHC, Upgrades.

## 1. Findings

The HL-LHC operating scenario is to level the instantaneous luminosity at $5 \times 10^{34}$ cm$^{-2}$s$^{-1}$ from a potential peak value of $1 \times 10^{35}$ cm$^{-2}$s$^{-1}$ at the beginning of fill, and to provide 250 fb$^{-1}$ per year for about 10 years. Operating the HL-LHC with 25 ns bunch crossing spacing at at $5 \times 10^{34}$ cm$^{-2}$s$^{-1}$ imply a pileup of up to 200 min-bias events/crossing. This is an order of magnitude greater than the LHC design luminosity ($10^{34}$) figure of 17 min-bias events per crossing and will degrade all occupancy-dependent trigger algorithms that rely on forms of isolation to identify electrons, muons, taus and missing energy signals. This requires a better performing trigger with additional information, such as tracking data, used to reduce the trigger rates against the much higher backgrounds. The size of regions sampled for trigger decisions will need to shrink to handle the increased backgrounds. Therefore, substantial improvements will be needed for the ATLAS[1] and CMS[2] Trigger systems for HL-LHC.

The addition of track information in L1 Trigger algorithms can provide significant improvements, particularly for lepton triggers, and helps with the overall rate problem, although hadronic triggers may still require higher bandwidth. L1 tracking triggers can sharpen the thresholds for lepton triggers and identify which objects (e.g. jets) are associated with the primary vertex. L1 tracking trigger calculations could include the provision of a pixel trigger in combination with calorimeter and muon trigger information with that of the tracking trigger to provide b-tagging and improve isolation and lepton identification. The tracking trigger reconstruction of tracks with good efficiency down to a $p_T$ of ~2.5 GeV or lower would be needed to provide lepton isolation.

The new trigger logic must operate in a short time envelope or latency unless substantial front-end electronics modifications are made in the ATLAS and CMS Detectors since these trigger calculations may require more latency than available in the present front-end electronics buffers.

Modification of the front-end electronics is also motivated by the need for a higher rate level-1 trigger.

The increased selectivity of the better performing L1 trigger and the increased complexity of the higher occupancy events will place a greater burden on the Higher Level Triggers, requiring a substantial increase in processing power.

## 2. Recommendations

Studies of the hardware capabilities and firmware technology to fully exploit advances in Field Programmable Gate Arrays (FPGAs), including architectures for the large increase in logic cells, DSPs, embedded linux and high speed and large number of serialization/deserialization stages.

Studies of high-speed optical link technologies, new compact high-density optical connectors and receivers to understand integration into trigger board designs.

Studies of commercial high speed electronics infrastructure systems such as the Advanced Telecommunications Computing Architecture (ATCA) and Micro-TCA derived from the Advanced Mezzanine Card standard to understand how to leverage high-speed star and mesh backplanes and the control technologies to provide the most effective infrastructure for trigger systems.

Studies of alternative processing units for dedicated higher-level trigger calculations such as Graphical Processor Units (GPUs), ARM and Xeon Phi.

Studies of Higher Level Trigger algorithm code to improve parallelization and thereby speed up CPU time and exploitation of different processor technologies.

## 3. Executive Summary

In order to reap the full physics harvest from the HL-LHC, it is necessary to maintain the physics acceptances of the key leptonic as well as hadronic and photon trigger objects such that the overall physics acceptances, especially for low-mass scale process like Higgs production, can be kept similar to those of the 2012 LHC running.

There are two strategies, which probably both need to be applied in order to achieve the above goal. The first one is the addition of a L1 tracking trigger for identification of tracks associated with calorimeter and muon trigger objects at L1. The second is a significant increase of L1 rate, L1 latency and HLT output rate.

Although the addition of a track trigger at L1 will help to maintain the rate requirements for leptonic trigger objects such as muons, electrons and possibly also taus at reasonable levels, only limited improvement is expected for photon triggers (except for track isolation) and hadronic trigger objects (except for requiring jets to have tracks pointing at the vertex). Therefore efficient triggering on photon and important hadronic objects at the HL-LHC may require increasing the L1 acceptance rate substantially beyond the present constraint of about 100 kHz, possibly as high as 1 MHz. Furthermore, a significant increase of the L1 latency the present 3-4 μsec would provide more time for tracking trigger calculations (including pixel tracking), for combination of calorimeter and muon trigger information with that of the tracking trigger and would provide additional flexibility which in turn could be used to facilitate the L1 decisions and thus to reduce the rate required on key trigger objects.

A source of trigger primitives not used in the current LHC L1 trigger systems are the strip and pixel trackers. Presently tracker information is added only in the HLT, where it effectively reduces rates and backgrounds. The strip tracker can provide information of four types: (1) the simple presence of a track match validates a calorimeter or muon trigger object, e.g. discriminating electrons from hadronic ($\pi^0 \rightarrow \gamma\gamma$) backgrounds in jets; (2) linking of precise tracker system tracks to muon system tracks improves precision on the $p_T$ measurement, sharpening thresholds in the muon trigger; (3) the degree of isolation of a e, γ, μ or τ candidate; and (4) the primary z-vertex location within the 30 cm luminous region derived from projecting tracks found in trigger layers, providing discrimination against pileup events in multiple object triggers, e.g. in lepton plus jet triggers. Jets inconsistent with the lepton vertex could be rejected. In the HLT, track matching to electron L1 objects reduces the rate by a factor of 10. A similar rejection factor is achieved for muons by adding tracker measurements for isolation and $p_T$. The goal is to use tracking information to accomplish a similar L1 rejection factor as is currently accomplished at the HLT level. A L1 pixel trigger offers the opportunity to tag secondary vertices for identification of b meson decays. A L1 pixel track trigger would also provide improved electron identification from track matching and isolation since the tracks are produced before there is a significant probability for conversion. Needless to say, the provision of strip or pixel L1 track trigger information has significant consequences for the design of the strip and pixel tracking detectors and their electronics. In addition, its full exploitation requires the L1 trigger systems to use the smallest possible resolution input trigger information from the calorimeter and muon systems and to process and process this information while maintaining this fine resolution for combination with the tracking trigger information. Provision of the finer resolution information from the calorimeter and muon detectors as well as the implications of providing trigger information for the pixel and strip trackers is covered in separate documents on the design of the upgrades for these detectors.

Another source of new information not widely exploited in L1 triggering is fast timing, e.g. in calorimeters. The CMS $PbWO_3$ crystal calorimeter has a resolution of 150 ps and other crystals (LYSO) have resolutions less than 100 ps. The timing of calorimeter pulses can be used to eliminate pile-up of overlapping interactions by distinguishing the source of the energy deposit between the hard scatter of interest and other interactions.

The hardware implementation of the Level-1 trigger upgrades uses high-bandwidth serial I/O links for data communication and large, modern field-programmable gate arrays (FPGAs) for sophisticated and fast algorithms. The latest developments and expectations for future FPGAs not only include significant increases in the number of logic gates available and high speed serial links, but also increases in the number of high-bandwidth serial links per device, more sophisticated and fast DSPs, embedded Linux (e.g. using Xilinx micro-blaze), and integration with high speed networking. Fast Tracking Trigger devices such as the ATLAS FTK use Associative Memories. The realization of the hardware also should exploit new modern standards deployed in industry where possible, such as the Advanced Telecommunications Architecture (TCA) for backplanes, which offers significantly more backplane bandwidth and flexibility (*i.e*. supports GbE, SATA/SAS, PCIe and SRIO) and provides for users to extend the backplane connectivity using the spare I/O available on each card.

The increase in L1 output rate from 100 kHz to possibly as high as 1 MHz requires higher bandwidth into the DAQ system and more CPU power in the HLT. The addition of a tracking trigger and more sophisticated algorithms at L1 means that the purity of the sample of events passing the L1 trigger is higher and that many of the algorithms heretofore used by the HLT are deployed in the L1 trigger, requiring a higher sophistication and complexity of algorithms used at the HLT. This would imply a greater CPU power than scaling with the L1 output rate. However, this is somewhat mitigated by the availability of the L1 Tracking Trigger primitives in the data immediately accessible by the HLT. Without a L1 tracking trigger, the opportunity to access most of the tracker information at the first levels of the HLT is limited by the CPU time to unpack and reconstruct the tracking data. This is significantly improved in the ATLAS experiment Phase 1 (pre-HL-LHC) upgrade with the ATLAS Fast Tracker (FTK)[3] project that will provide quick access to tracking information in the ATLAS Level-2 trigger. For Phase 2 (HL-LHC), the addition of the L1 tracking trigger means that the results from the L1 track reconstruction can be immediately used without the overhead of tracking data unpacking and reconstruction. This will further aid the HLT in reducing the CPU load associated with use of tracking information in the initial algorithms.

In networking technology there are developments in Ethernet as well as High Performance Computing (HPC) interconnects. Network Interface Cards (NICs) are now available for 40 GbE. It is reasonable to expect that by the end of this decade links of 100 Gbps (based on 4 lanes of 25 Gbps) are available at reasonable cost. For the event building, assuming a 10 MB event size at 1 MHz, the readout requires 100 Tbps bandwidth (assuming 80% efficiency), which should be achievable considering that switches today provide 32 Tbps.

If the HLT is based on commodity CPUs, the trend for dual-CPU servers shows a 20% cost improvement per year (1). This would result in a about a factor 10 cost/performance improvement in 10 years. In this scenario there is little distinction between the HLT and the first stage of offline computing, which offers opportunities for moving code and functionality between the two. An example would be for either ATLAS or CMS to share their HLT Farm and Tier-0 computing resources so that at the beginning of a fill when the luminosity might be higher, the HLT would also use part of the Tier-0 resources. Later in the fill when the luminosity had dropped and between fills, the Tier-0 would use the HLT resources.

The evolution of the computing market towards different computing platforms offers an opportunity to achieve substantial gains in HLT processing power at the price of adapting code to the new hardware. Examples include Graphical Processor Units (GPUs), such as the NVIDIA Tesla and GeForce (presently used by the Alice Collaboration[4]), ARM processors (used in smartphones and now exceeding the deployed base of x86 architecture processors) and the Intel Xeon Phi coprocessor. As is the case for enhancing computing performance, the adaptation of code may require increasing parallelization. The topic of optimizing code and exploiting new CPU technologies is covered in more detail in the documents on computing.